\documentclass[twocolumn,prd,preprintnumbers,amsmath,amssymb,superscriptaddress]{revtex4-1}

\usepackage{amsmath,amssymb,graphicx,float,colordvi,url,wrapfig}
\usepackage[dvips]{color}
\usepackage{wrapfig}

\allowdisplaybreaks
\begin{document}




\title{Complex Langevin studies of the dynamical compactification of extra dimensions in the Euclidean IKKT matrix model
}

\author{Konstantinos N.~{\sc Anagnostopoulos}}
\email[]{konstant@mail.ntua.gr}
\affiliation{
{\it National Technical University of Athens, Zografou Campus, GR-15780 Athens, Greece}}

\author{Takehiro~{\sc Azuma}}
\email[]{azuma@mpg.setsunan.ac.jp}
\affiliation{
{\it Setsunan University,
17-8 Ikeda Nakamachi, Neyagawa, Osaka, 572-8508, Japan }}

\author{Yuta~{\sc Ito}}
\email[]{y-itou@tokuyama.ac.jp}
\affiliation{{\it National Institute of Technology, Tokuyama College, Gakuendai, Shunan, Yamaguchi 745-8585, Japan}}

\author{Jun~{\sc Nishimura}}
\email[]{jnishi@post.kek.jp}
\affiliation{{\it High Energy Accelerator Research Organization (KEK), 1-1 Oho, Tsukuba, Ibaraki 305-0801, Japan }}
\affiliation{{\it Graduate University for Advanced Studies (SOKENDAI),
1-1 Oho, Tsukuba, Ibaraki 305-0801, Japan}}

\author{Toshiyuki~{\sc Okubo}}
\email[]{tokubo@meijo-u.ac.jp}
\affiliation{{\it Faculty of Science and Technology, Meijo University,
Nagoya, 468-8502, Japan}}

\author{Stratos Kovalkov~{\sc Papadoudis}}
\email[]{sp10018@central.ntua.gr}
\affiliation{
{\it National Technical University of Athens, Zografou Campus, GR-15780 Athens, Greece}}

\begin{abstract}
The type IIB matrix model, also known as the IKKT matrix model, is a promising candidate for a nonperturbative formulation of superstring theory. 
In this talk we study the Euclidean version of the IKKT matrix model, which has a ``sign problem'' due to the Pfaffian coming from integrating out the fermionic degrees of freedom.  To study the spontaneous breaking of the SO(10) rotational symmetry, we apply the Complex Langevin Method (CLM) to the Euclidean IKKT matrix model. We conclude that the SO(10) symmetry is broken to SO(3), in agreement with the previous studies by the Gaussian Expansion Method (GEM). We also apply the GEM to the deformed model and find consistency with the CLM result. These are proceedings of Takehiro Azuma's talk at Asia-Pacific Symposium for Lattice Field Theory (APLAT 2020) on August 4-7, 2020, based on the paper \cite{2002_07410}.
\end{abstract}

\maketitle


\section{Introduction}
Large-$N$ reduced models have been proposed as the non-perturbative definition of superstring theory. In particular, the type IIB matrix model, also known as the IKKT matrix model \cite{9612115}, is regarded as one of the most promising approaches. The theory is formally defined by the dimensional reduction of the ten-dimensional ${\cal N}=1$ super-Yang-Mills theory to zero dimensions. We interpret the eigenvalues of the bosonic matrices as the spacetime coordinate, and the spacetime is dynamically generated from the matrices' degrees of freedom. Superstring theory is well-defined in the ten-dimensional spacetime, and it is an important question how our four-dimensional spacetime emerges dynamically.

The Euclidean version of the IKKT matrix model is obtained after a Wick rotation of the temporal direction. It has a manifest SO(10) rotational symmetry, whose spontaneous breaking implies the dynamical generation of the lower-dimensional spacetime. In the Euclidean version, it has been known that the spontaneous symmetry breaking (SSB) of SO(10) is not realized in the phase-quenched model, and that the complex phase of the Pfaffian that comes from integrating out the fermionic degrees of freedom plays an important role in the SSB of SO(10) \cite{9811220,0003208,0003223,0005147,1306_6135,1509_05079}. On the other hand, it is difficult to numerically study the systems with complex phase, due to the so-called ``complex action problem'', or ``sign system''. If we are to study the SSB of SO(10), we need to overcome the ``sign problem''.

Various approaches to the ``sign problem'' have been so far proposed. One of the promising approaches to the ``sign problem'' is the ``Complex Langevin Method'' (CLM) \cite{Parisi:1984cs,Klauder:1983sp}, which attempts to define a stochastic process for the complexified variables. Recently, the CLM has gained enormous attention because the condition for the equivalence to the original path integral has been clarified \cite{0912_3360,1101_3270,1211_3709,1504_08359,1508_02377,1604_07717,1606_07627}. 
The CLM has been previously applied to the toy models to capture the rotational symmetry breaking in the Euclidean IKKT matrix model \cite{1609_04501,1712_07562}. In  \cite{1712_07562}, five of the authors (K.N.A., T.A., Y.I., J.N. and S.K.P.) have investigated the six-dimensional version of the Euclidean IKKT matrix model. Similarly to the IKKT matrix model, the six-dimensional version also has a ``sign problem'' from the determinant which is obtained by integrating out the fermion. Using the CLM, it has been shown that the SO(6) rotational symmetry is broken to SO(3), as suggested by Gaussian Expansion Method (GEM) \cite{1007_0883,1108_1293}.

In this talk, we apply the CLM to the Euclidean IKKT matrix model, extending the study of the six-dimensional version \cite{1712_07562}. This is numerically much more involved than the six-dimensional version, since the size of the gamma matrices after Weyl projection increases from 4 to 16. Also, the finite-$N$ effects become severer, which requires simulations at large $N$ to make sensible large-$N$ extrapolations. This makes the extrapolation with respect to the deformation parameters, which were used for the CLM of the toy models \cite{1609_04501,1712_07562}, more difficult. This leads us to study the model with a mass deformation of the fermion using the GEM, which is compared with the CLM result. 

These proceedings are organized as follows. In Sec. \ref{Sec_EIKKT}, we introduce the Euclidean version of the IKKT matrix model, which we focus on. In Sec. \ref{Sec_CLM}, we apply the CLM to the IKKT matrix model. In Sec. \ref{Sec_GEM}, we present the result of the GEM applied to the deformed model, which is well compared with the CLM result. Sec. \ref{Sec_Conclu} is devoted to a summary and discussions.

\section{Euclidean version of the IKKT matrix model} \label{Sec_EIKKT}
The action $S$ of the IKKT matrix model  \cite{9612115} is given by
\begin{eqnarray}
 S &=& S_{\textrm{b}} + S_{\textrm{f}}, \textrm{ where } \label{IKKT_action} \\
 S_{\textrm{b}} &=& - \frac{N}{4} \textrm{tr} [A_{\mu}, A_{\nu}] [A^{\mu}, A^{\nu}], \label{IKKT_boson} \\
 S_{\textrm{f}} &=& - \frac{N}{2}  \textrm{tr} \left( {\bar\psi}_{\alpha} (\Gamma^{\mu})_{\alpha \beta} [A_{\mu}, \psi_{\beta}] \right). \label{IKKT_fermion}
\end{eqnarray}
The bosons $A_{\mu}$ ($\mu =1,2, \cdots, 10$) are $N \times N$ traceless Hermitian matrices, and the Majorana-Weyl spinors $\psi_{\alpha}$ ($\alpha =1,2, \cdots, 16$) are  $N \times N$ traceless matrices with Grassmann entries. $\Gamma^{\mu}$ are the $16 \times 16$ gamma matrices after Weyl projection. $\ {\bar \psi} = \psi {\cal C}$, with ${\cal C}$ being the $16 \times 16$ charge conjugation matrix. In the Euclidean version, the indices are contracted by $ \delta_{\mu \nu} = (1,1,1,\cdots,1)$, and the partition function is
\begin{eqnarray}
 \hspace*{-6mm} Z = \int dA d\psi e^{-S} = \int dA e^{-S_{\textrm{b}}} \textrm{Pf } {\cal M} = \int dA e^{-S_{\textrm{eff}}}.  \label{EIKKT_partition}
\end{eqnarray}
The effective action $S_{\textrm{eff}}$ is defined as
\begin{eqnarray}
 S_{\textrm{eff}} = S_{\textrm{b}} - \log \textrm{Pf } {\cal M}. \label{Seff_definition}
\end{eqnarray}
${\cal M}$ is a $16 (N^2-1) \times 16(N^2-1)$ anti-symmetric matrix, which represents a linear transformation
\begin{eqnarray}
 \psi_{\alpha} \to ({\cal M} \psi)_{\alpha} = ({\cal C} \Gamma^{\mu})_{\alpha \beta} [A_{\mu}, \psi_{\beta}], \label{M_definition}
\end{eqnarray}
acting on the linear space of traceless complex $N \times N$ matrices $\psi_{\alpha}$.
In the Euclidean version, the partition function is finite without any cutoff \cite{9803117,0103159}. However, the Pfaffian $\textrm{Pf } {\cal M}$ is complex in general and we face a severe sign problem. We define its phase $\Gamma$ as $ \textrm{Pf } {\cal M} = |\textrm{Pf } {\cal M}| e^{i \Gamma}$. It has been shown that in the phase-quenched model, in which $ \textrm{Pf } {\cal M}$ is replaced with $ |\textrm{Pf } {\cal M}|$, the SSB of the SO(10) rotational symmetry does not occur \cite{0003208,0005147,1306_6135,1509_05079}.

The SSB of the SO(10) symmetry has been studied via the GEM, and it has turned out that SO(10) is spontaneously broken to SO(3)  \cite{1007_0883,1108_1293}. We consider the $10 \times 10$ ``moment of inertia tensor''
\begin{eqnarray}
 T_{\mu \nu} = \frac{1}{N} \textrm{tr} (A_{\mu} A_{\nu}). \label{t_munuE}
\end{eqnarray}
We define its 10 eigenvalues as $\lambda_{\mu}$ with the ordering $ \lambda_1 > \lambda_2 > \cdots > \lambda_{10}$. In the SO($d$) vacuum, the V.E.V.'s $ \langle \lambda_1 \rangle, \cdots, \langle \lambda_d \rangle$ grow and the V.E.V.'s $ \langle \lambda_{d+1} \rangle, \cdots, \langle \lambda_{10} \rangle$ shrink in the large-$N$ limit. The results of the studies of the SO$(d)$ symmetric vacua for $2 \leq d \leq 7$ are summarized as follows:\\
\noindent 1. The extent of the shrunken directions $r = \lim_{N\to \infty} \sqrt{\lambda_n}$ ($n=d+1, \cdots,10$) is $r^2 \simeq 0.155$, which does not depend on $d$ (universal compactification scale).\\
\noindent 2. The ten-dimensional volume of the Euclidean spacetime does not depend on $d$, except for $d=2$ (constant volume property). For the extent of the extended directions $R = \lim_{N \to \infty} \sqrt{\lambda_n}$ ($n=1,2,\cdots,d$), the volume is $V = R^d r^{10-d} = l^{10}$, with $l^2 \simeq 0.383$.\\
\noindent 3. The free energy takes the minimum value at $d=3$, which suggests the dynamical emergence of {\it three}-dimensional spacetime.\\
The Euclidean version has been studied by Monte Carlo simulation using the factorization method \cite{1306_6135,1509_05079}. These works provided strong numerical evidence for the realization of SSB of the SO(10) rotational symmetry, but they were not able to determine the precise SSB pattern accurately enough.

\section{Complex Langevin studies of the IKKT matrix model} \label{Sec_CLM}
In this section, we apply the CLM to the effective action of the IKKT matrix model $S_{\textrm{eff}}$ defined by (\ref{Seff_definition}). In the CLM, we extend the degrees of freedom of $A_{\mu}$ from the Hermitian traceless matrices to the general complex traceless matrices. The CLM consists of solving the following Langevin equation 
\begin{eqnarray}
 \hspace*{-5mm} \frac{d A_{\mu}(t)_{ij}}{dt} &=& - \frac{\partial S_{\textrm{eff}}}{\partial A_{\mu}(t)_{ji}} + \eta_{\mu} (t)_{ij}, \textrm{ where } \label{CLM_eq_cont} \\
 \hspace*{-5mm} \frac{\partial S_{\textrm{eff}}}{\partial A_{\mu}(t)_{ji}} &=&   \frac{\partial S_{\textrm{b}}}{\partial A_{\mu}(t)_{ji}} - \frac{1}{2} \textrm{Tr} \left( \frac{\partial {\cal M}}{\partial A_{\mu}(t)_{ji}} {\cal M}^{-1} \right). \label{drift_term}
\end{eqnarray}
Tr is the trace with respect to the $16 (N^2-1) \times 16(N^2-1)$ matrix.  We call (\ref{drift_term}) the ``drift term''.  $t$ is a fictitious time, which should not be confused with the real time. 
$\eta_{\mu} (t)_{ij}$ is the Hermitian white noise that follows the probability distribution $ \propto \exp \left( - \frac{1}{4} \int \textrm{tr} \eta^2 (t) dt \right)$. 
This is independent for different times $t, t'$. The white noise is rendered traceless as $ \eta(t)_{ii} \to \eta(t)_{ii} - \frac{1}{N} \textrm{tr} \eta(t)$. The expectation value of an observable is evaluated as
\begin{eqnarray}
 \langle {\cal O} [A_{\mu}] \rangle = \frac{1}{T} \int^{t_0+T}_{t_0} {\cal O} [A_{\mu} (t)] dt. \label{VEV_CLM}
\end{eqnarray}
$t_0$ is the thermalization time, and $T$ is sufficiently large to obtain good statistics. The holomorphy of the observable ${\cal O}$ is important for the proof of the validity of  (\ref{VEV_CLM}) \cite{0912_3360,1101_3270,1606_07627}. The Langevin equation (\ref{CLM_eq_cont}) is put on a computer by the discretization
{\small
\begin{eqnarray}
 (A_{\mu})(t+\Delta t)_{ij} = (A_{\mu})(t)_{ij} - (\Delta t) \frac{\partial S_{\textrm{eff}}}{\partial A_{\mu}(t)_{ji}} + \sqrt{(\Delta t)} \eta_{\mu} (t)_{ij}. \nonumber \\ \label{CLM_eq_discre}
\end{eqnarray}}
The term $\sqrt{(\Delta t)}$ stems from the normalization of $\eta_{\mu}(t)$, so that it follows the probability distribution $ \propto \exp \left( - \frac{1}{4} \sum_t \textrm{tr} \eta^2 (t)  \right)$.

We cannot extract a reliable result equivalent to the path integral from the CLM when we encounter the following two problems: One is the ``excursion problem'', which occurs when $A_{\mu}$ is too far from Hermitian. The other is the ``singular drift problem'', which occurs when the drift term (\ref{drift_term}) becomes large due to the accumulation of some of the eigenvalues of ${\cal M}$ close to zero. In order to justify the CLM, the probability distribution of the ``drift norm''
\begin{eqnarray}
 u = \sqrt{\frac{1}{10N^3} \sum_{\mu=1}^{10} \sum_{i,j=1}^{N} \left|  \frac{\partial S_{\textrm{eff}}}{\partial (A_{\mu})_{ji}} \right|^2 }, \label{drift_norm_def}
\end{eqnarray}
which is measured during the complex Langevin simulation, should fall off exponentially or faster. If we look at the ``drift term'', we get the drift of the CLM \cite{1606_07627}.

In order to avoid the ``excursion problem'', we use the technique named ``gauge cooling'' \cite{1211_3709,1604_07717}, to keep $A_{\mu}$ closer to Hermitian matrices. This consists of minimizing the ``Hermiticity norm'' defined by ${\cal N}_{\textrm{H}} = - \frac{1}{10N} \sum_{\mu=1}^{10} \textrm{tr} \{ (A_{\mu} - A_{\mu}^{\dagger})^2 \}$
at each step in solving the discretized Langevin equation (\ref{CLM_eq_discre}).

In applying the CLM to the IKKT matrix model, we add the following two mass terms to the action $S$ as defined in (\ref{IKKT_action}) \cite{1609_04501,1712_07562}:
\begin{eqnarray}
 \Delta S_{\textrm{b}} &=& \frac{N}{2} \varepsilon \sum_{\mu=1}^{10} m_{\mu} \textrm{tr} (A_{\mu})^2, \label{boson_mass} \\
 \Delta S_{\textrm{f}} &=& -im_{\textrm{f}} \frac{N}{2} \textrm{tr} (\psi_{\alpha} ({\cal C} \Gamma_8 \Gamma_9^{\dag} \Gamma_{10} )_{\alpha \beta} \psi_{\beta}). \label{fermion_mass}
\end{eqnarray}
In order to probe the SSB, we break the SO(10) symmetry explicitly by adding the bosonic mass term (\ref{boson_mass}). Here, $m_{\mu}$ satisfies $0<m_1 \leq \cdots \leq m_{10}$. We consider the following order parameter for the SSB of SO(10):
\begin{eqnarray}
 \lambda_{\mu} = \frac{1}{N} \textrm{tr} (A_{\mu})^2, \ \ (\mu=1,2,\cdots,10). \label{SSB_order_para}
\end{eqnarray}
Here, there is no summation of $\mu$. We consider (\ref{SSB_order_para}) instead of the eigenvalues of $T_{\mu \nu}$ defined by (\ref{t_munuE}), to avoid the subtleties in the holomorphy of the observables. We take the $\varepsilon \to 0$ limit {\it after} taking the large-$N$ limit.

The mass term (\ref{fermion_mass}) is added to avoid the singular drift problem coming from the near-zero eigenvalues of ${\cal M}$, by shifting the eigenvalue distribution of ${\cal M}$ on the complex plane away from the origin. The mass term  (\ref{fermion_mass}) breaks the SO(10) symmetry to SO(7) $\times$ SO(3). We study whether SO(7) is broken to smaller subgroups as we vary $m_{\textrm{f}}$, and discuss what happens in the $m_{\textrm{f}} \to 0$ limit. The $m_{\textrm{f}} \to +\infty$ limit is the bosonic model, since the fermionic degrees of freedom decouple. It is known that there is no SSB of SO(10) in the bosonic model \cite{9811220}. In our simulations, we choose the range of the parameters $(m_{\textrm{f}}, \varepsilon)$ for each $N$, so that the probability distribution of the ``drift norm'' (\ref{drift_norm_def}) falls off exponentially or faster.

\begin{figure*}
\centering
\includegraphics[width=0.401\textwidth]{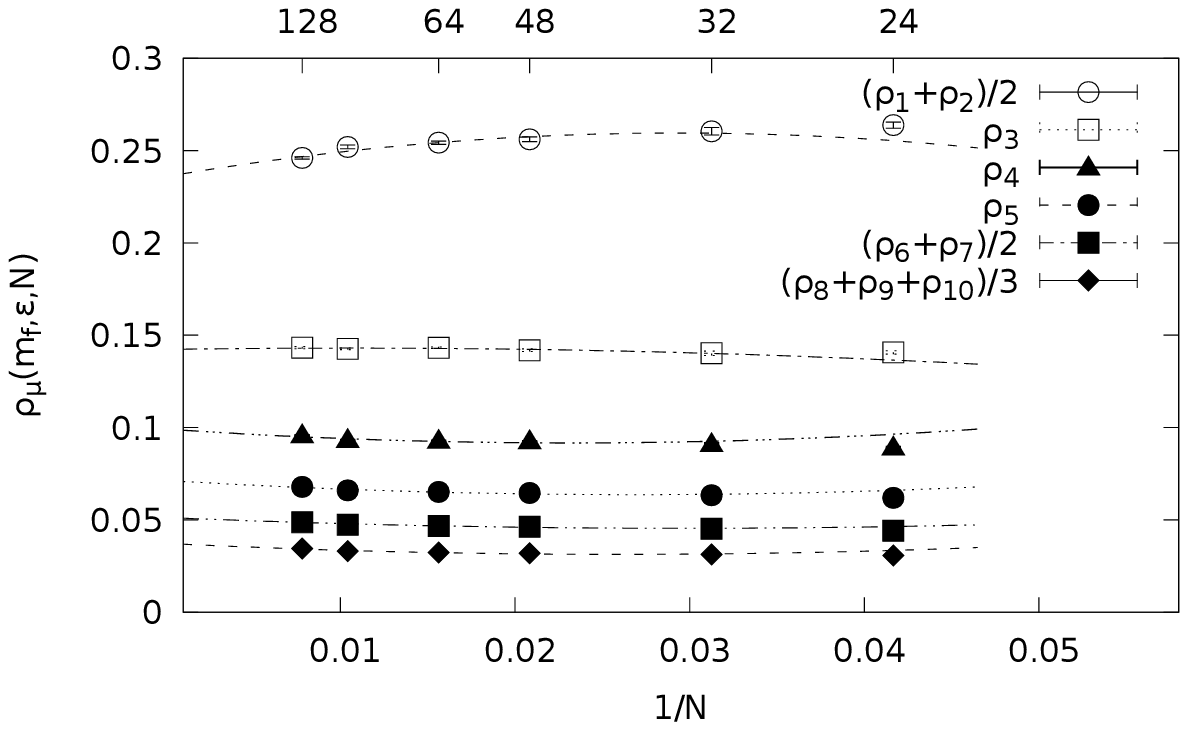} 
\includegraphics[width=0.401\textwidth]{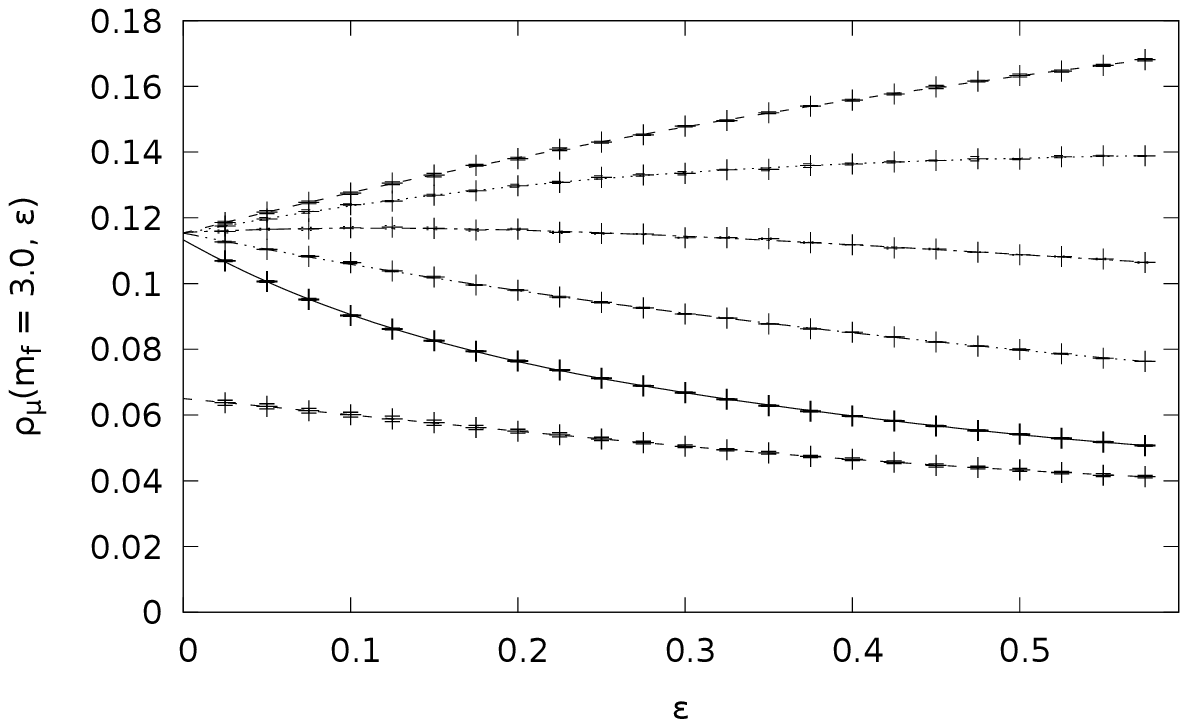} 
\includegraphics[width=0.401\textwidth]{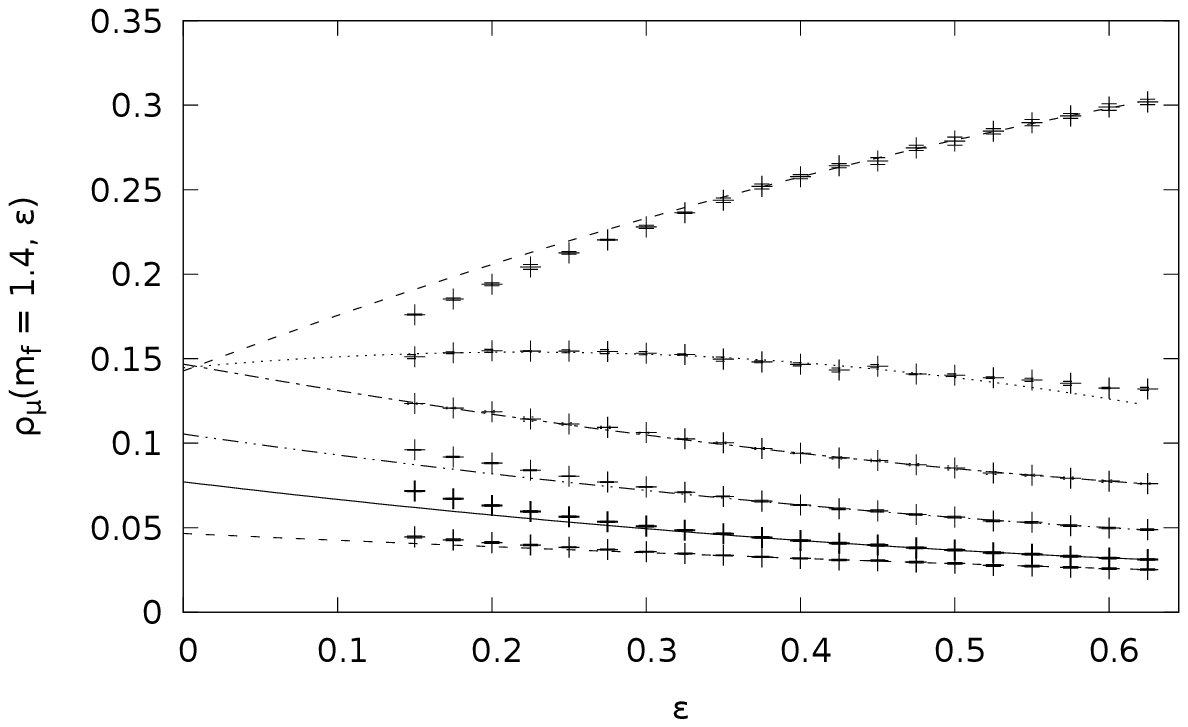}
\includegraphics[width=0.401\textwidth]{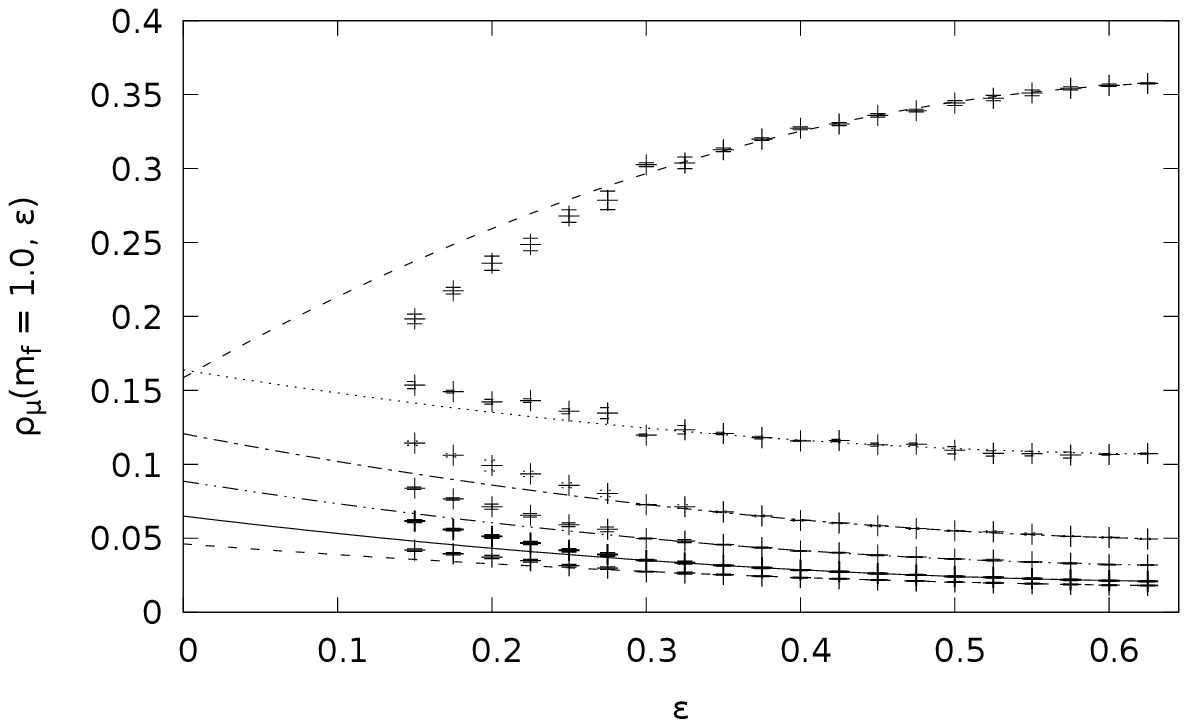}
\includegraphics[width=0.401\textwidth]{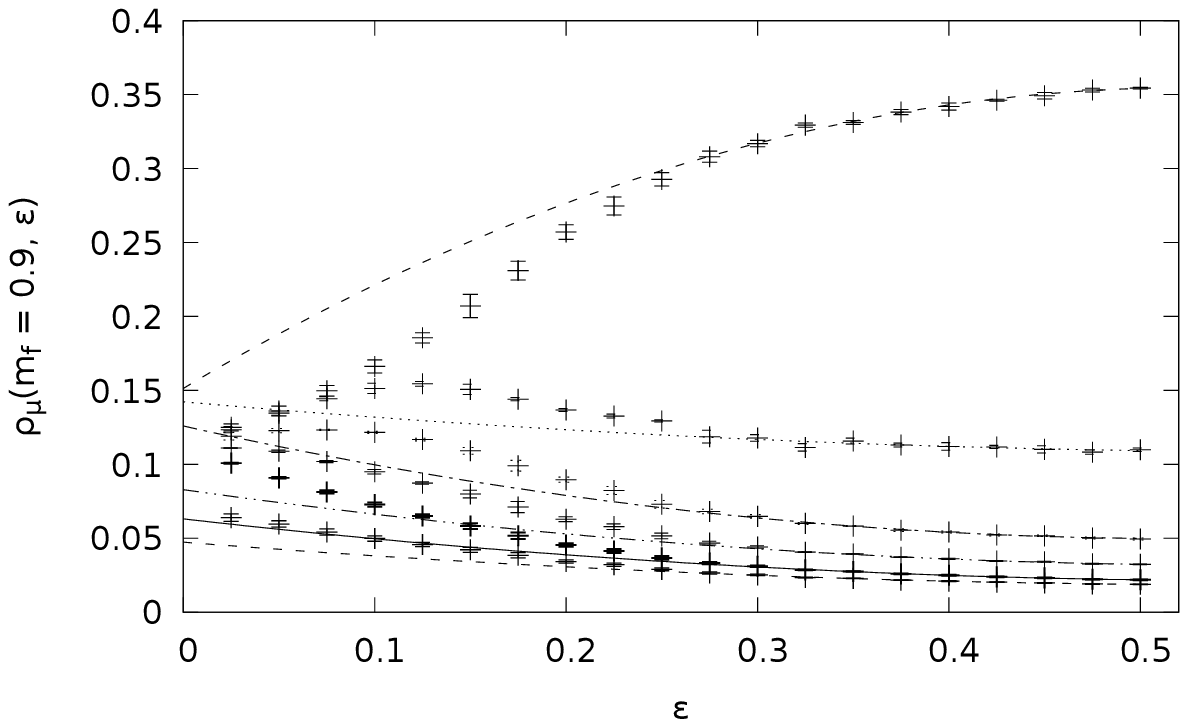}
\includegraphics[width=0.401\textwidth]{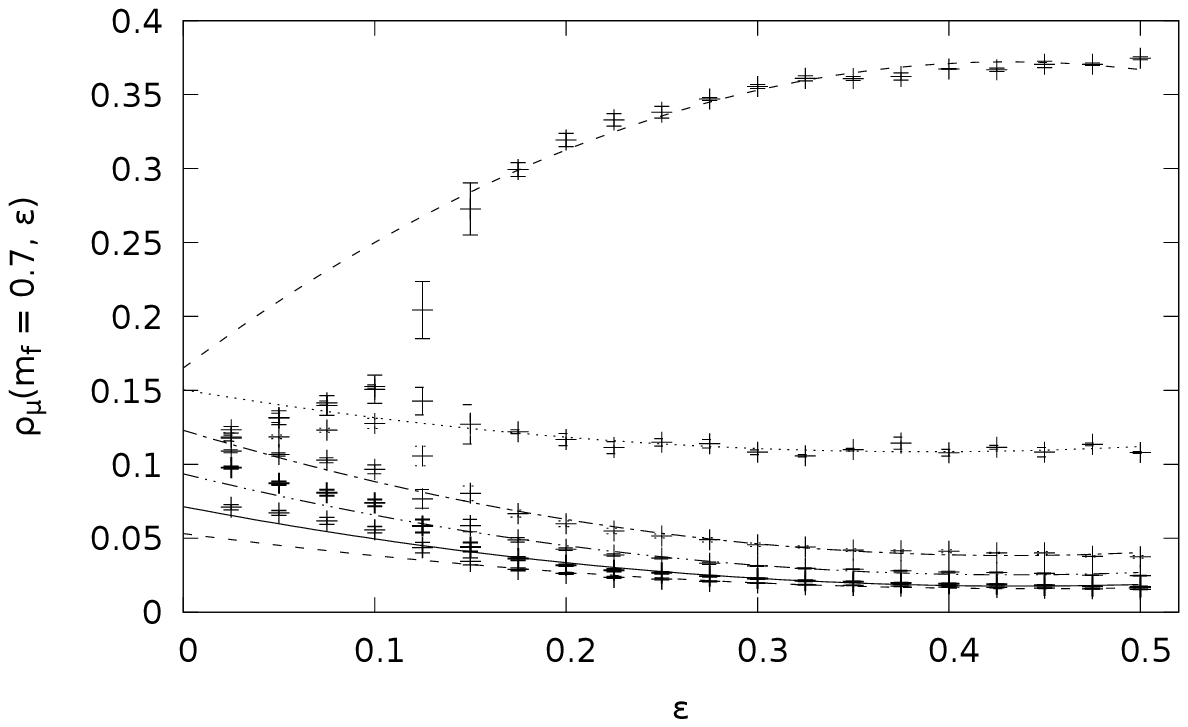}
\vspace*{-3mm}
    \caption{(Top Left) The large-$N$ extrapolation of $\rho_{\mu} (m_{\textrm{f}}, \varepsilon, N)$ for $ m_{\textrm{f}} = 1.0, \varepsilon = 0.2$, with $m_{\mu}$ given by (\ref{m_mu2}). The $\rho_{\mu} (m_{\textrm{f}}, \varepsilon, N)$ are averaged for $\mu=1,2$, $\mu=6,7$ and $\mu=8,9,10$ in order to increase statistics. We make a quadratic fit with respect to $\frac{1}{N}$. \protect\linebreak
(The Rest) $ \rho_{\mu} (m_{\textrm{f}}, \varepsilon)$ after taking the large-$N$ limit are plotted against $ \varepsilon$ for $ m_{\textrm{f}}=3.0$ (Top Right), $ m_{\textrm{f}}=1.4$ (Middle Left), $ m_{\textrm{f}}=1.0$ (Middle Right), $ m_{\textrm{f}}=0.9$ (Bottom Left) and $ m_{\textrm{f}}=0.7$ (Bottom Right). Quartic and Quadratic fits are performed at  $ m_{\textrm{f}}=3.0$ and  $ m_{\textrm{f}}\leq 1.4$, respectively. The curves from top to bottom are $ \frac{1}{3} (\rho_1+\rho_2+\rho_3), \rho_4, \rho_5, \rho_6, \rho_7, \frac{1}{3} (\rho_8+\rho_9+\rho_{10})$ for $ m_{\textrm{f}}=3.0$, and $ \frac{1}{2} (\rho_1+\rho_2), \rho_3, \rho_4, \rho_5, \frac{1}{2} (\rho_6+\rho_7), \frac{1}{3} (\rho_8+\rho_9+\rho_{10})$ for $ m_{\textrm{f}}\leq 1.4$, respectively.}
\vspace*{-5mm}
\label{result_CLM}
\end{figure*}

We apply the CLM to the action, with the original model (\ref{IKKT_action}) deformed by (\ref{boson_mass}) and (\ref{fermion_mass}):
\begin{eqnarray}
 S' = S + \Delta S_{\textrm{b}} + \Delta S_{\textrm{f}}. \label{deformed_IKKT}
\end{eqnarray} 
The effective action (\ref{Seff_definition}) is modified accordingly. To make the $\varepsilon \to 0$ limit sensible, we need to choose $m_{\mu}$ carefully. Here, we take $m_{\mu}$ as
\begin{eqnarray}
 m_{\mu} &=& (0.5, 0.5, 0.5, 1,2,4,8,8,8,8) \ \  (m_{\textrm{f}} = 3.0), \label{m_mu1} \\
 m_{\mu} &=& (0.5, 0.5, 1,2,4,8,8,8,8,8) \ \  (m_{\textrm{f}} \leq 1.4),  \label{m_mu2} 
\end{eqnarray}
so that we can distinguish the SO$(d)$ vacua with $d=3,4,5,6,7$ at $m_{\textrm{f}}=3.0$, and $d=2,3,4,7$ at $m_{\textrm{f}}\leq 1.4$, respectively. First, we compute the ratio for finite $N$:
\begin{eqnarray}
 \rho_{\mu} (m_{\textrm{f}}, \varepsilon, N) = \frac{\langle \lambda_{\mu} \rangle_{m_{\textrm{f}}, \varepsilon, N}}{\sum_{\nu=1}^{10} \langle \lambda_{\nu} \rangle_{m_{\textrm{f}} , \varepsilon, N}}, \label{rho_finiteN}
\end{eqnarray}
where $\langle \lambda_{\mu} \rangle_{m_{\textrm{f}} , \varepsilon, N}$ is the V.E.V. of the observable (\ref{SSB_order_para}) with respect to the deformed action (\ref{deformed_IKKT}) at finite $N$. Then, we make a large-$N$ extrapolation
\begin{eqnarray}
 \rho_{\mu} (m_{\textrm{f}}, \varepsilon) = \lim_{N \to +\infty} \rho_{\mu} (m_{\textrm{f}}, \varepsilon, N).  \label{rho_largeN}
\end{eqnarray}
An example of this procedure is shown in FIG. \ref{result_CLM}. The Euclidean IKKT matrix model suffers so severe finite-$N$ effects that it requires a quadratic fit with respect to $\frac{1}{N}$.

After taking the large-$N$ limit, we plot $\rho_{\mu} (m_{\textrm{f}}, \varepsilon)$ against $\varepsilon$ in FIG. \ref{result_CLM}. Some of the small-$\varepsilon$ points are excluded from the fitting, due to the difficulty in overcoming the finite-$N$ effects. As we read off the $\varepsilon \to 0$ limit from the fitting, we see that at $m_{\textrm{f}}=3.0$ the SO(7) symmetry is unbroken. As we gradually lower $m_{\textrm{f}}$, we observe the SSB of SO(7) to SO(4) at $m_{\textrm{f}}=1.4$, and to SO(3) at $m_{\textrm{f}}=0.7,0.9,1.0$, respectively. This is consistent with the GEM result for the undeformed model (\ref{IKKT_action}), in which the SO(3) vacuum has the smallest free energy \cite{1108_1293}.

\section{Comparison with the GEM result} \label{Sec_GEM}
In this section, we present the result of the GEM for the action $S'' = S + \Delta S_{\textrm{f}}$, with the original model (\ref{IKKT_action}) deformed only by (\ref{fermion_mass}). The basic idea of the GEM is to rewrite the action $S''$ as
\begin{eqnarray}
 S'' &=& S_0 + (S''-S_0), \textrm{ where } \label{GEM_idea} \\
 S_0 &=& \frac{N}{2} \sum_{\mu=1}^{10} M_{\mu} \textrm{tr} (A_{\mu})^2 + \frac{N}{2} \sum_{\alpha, \beta=1}^{16} {\cal A}_{\alpha \beta} \textrm{tr} (\psi_{\alpha} \psi_{\beta}), \label{classic_GEM}  \\ 
 {\cal A}_{\alpha \beta} &=& - i m_{\textrm{f}} ({\cal C} \Gamma_8 \Gamma_9^{\dag} \Gamma_{10} )_{\alpha \beta} + \sum_{\mu,\nu,\rho=1}^{10} \frac{i}{3!} m_{\mu \nu \rho} ({\cal C} \Gamma_{\mu} \Gamma_{\nu}^{\dag} \Gamma_{\rho})_{\alpha \beta}.  \nonumber
\end{eqnarray}
$S_0$ and $(S''-S_0)$ are regarded as the ``classical action'' and the ``one-loop counter term'', respectively. We choose the ordering of the parameter $M_{\mu}$ as $0<M_1 \leq \cdots \leq M_{10}$. The other parameter $m_{\mu \nu \rho}$ is a totally anti-symmetric 3-form. 
We focus on the two cases, in which we impose SO$(d) \times Z_3$ with $d=6$ and $d=7$, where $Z_3$ represents the group of cyclic permutation of the 8th, 9th and 10th directions. In the following, we call this ``the SO$(d)$ ansatz''.

\begin{figure*}
\centering
\includegraphics[width=0.360\textwidth]{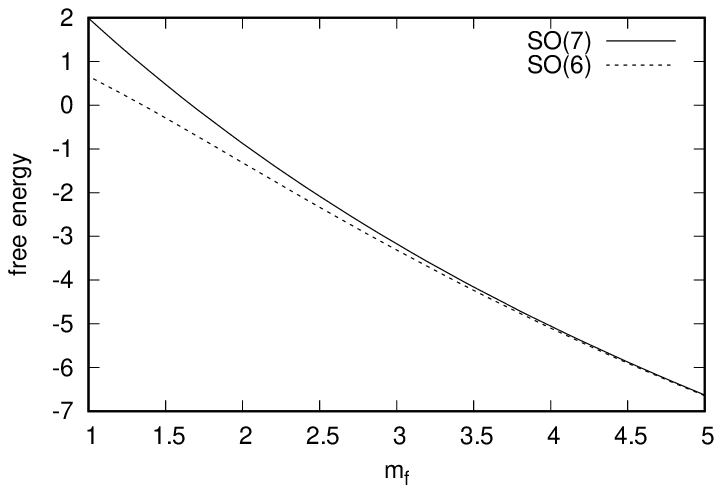} 
\includegraphics[width=0.400\textwidth]{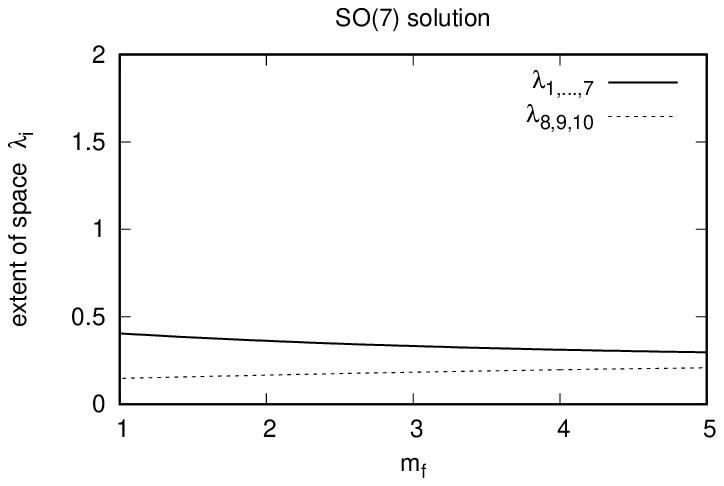} 
    \caption{(Left) The free energy calculated up to three loops for the solutions for SO(7) and SO(6) ansatzes are plotted against $m_{\textrm{f}}$. 
    (Right) The extent of space (\ref{SSB_order_para}) for the SO(7) ansatz is plotted against $m_{\textrm{f}}$. Here, we plot $ \lambda_{1}=\cdots =\lambda_7$ and $ \lambda_{8}= \lambda_9 =\lambda_{10}$.}
\vspace*{-3mm}    
\label{result_GEM}
\end{figure*}

In FIG. \ref{result_GEM} (Left) we plot the free energy calculated up to the three loops for the solutions found with the SO(7) and SO(6) ansatze against $m_{\textrm{f}}$. As we decrease $m_{\textrm{f}}$, we see a clear tendency that the SO(6) symmetric vacuum is favored. And the free energy for SO(7) and SO(6) is degenerate for large $m_{\textrm{f}}$. Also in FIG. \ref{result_GEM} (Right) we plot the extent of space $\lambda_{\mu}$ against $m_{\textrm{f}}$ for the SO(7) ansatz. At $m_{\textrm{f}}=3.0$, we have $\lambda_1 =\cdots \lambda_7 = 0.333, \ \ \lambda_8=\lambda_9=\lambda_{10} = 0.184$, which amounts to
\begin{eqnarray}
 \textrm{(GEM): }  \rho_1 = \cdots =  \rho_7 = 0.116, \ \ \rho_8 =\rho_9 =\rho_{10} = 0.064, \nonumber
\end{eqnarray}
where $ \rho_{\mu} = \frac{\lambda_{\mu}}{\sum_{\nu=1}^{10} \lambda_{\nu}}$. This is in impressive agreement with the CLM result 
\begin{eqnarray}
 \textrm{(CLM): }  \rho_1 = \cdots =  \rho_7 = 0.115, \ \  \frac{\rho_8 +\rho_9 +\rho_{10}}{3} = 0.065, \nonumber
\end{eqnarray}
which is read off from FIG. \ref{result_CLM} (Top-Right).

\section{Conclusion and outlook} \label{Sec_Conclu}
In this talk, we have studied the Euclidean version of the IKKT matrix model via the CLM, to elucidate how the spacetime is dynamically generated in superstring theory. Similarly to the previous work on the six-dimensional version \cite{1712_07562}, we have used the deformation technique to overcome the singular-drift problem. We have found that as we decrease $m_{\textrm{f}}$, the lower-dimensional spacetime is chosen. At $m_{\textrm{f}} \leq 1.0$, the SO(10) symmetry is spontaneously broken to SO(3). Although it is difficult to make a quantitatively sensible $m_{\textrm{f}} \to 0$ extrapolation of the ratio $\rho_{\mu}$, we conclude that the SO(3) vacuum is chosen in the undeformed model ($m_{\textrm{f}} = 0$). Also, we have studied the mass deformed model using the GEM, and compared the free energy for the SO(7) and SO(6) ansatze. Here again, as we decrease  $m_{\textrm{f}}$, the ansatze for lower-dimensional spacetime are energetically favored. At $\displaystyle m_{\textrm{f}}=3.0$, where the SO(7) vacuum is realized, we have seen a quantitative agreement between the CLM and GEM results of the spacetime extent.

The CLM can be applied to many interesting systems with the sign problem. In  \cite{1904_05919}, an attempt has been made to apply the CLM to the Lorentzian version of the IKKT matrix model, where the complex phase comes from $ e^{i S_{\textrm{b}}}$. It was found that for a certain deformed bosonic model the emergent three-dimensional expanding space has a clear departure from the fuzzy-sphere-like Pauli-matrix structure. We hope to pursue this direction and elucidate the structure of the  expanding space as suggested by superstring theory. Also, to study the nature of the expansion of the universe, we need to simulate the model at larger $N$. The impact on the fermionic degrees of freedom should be studied. We hope to report on more analysis in future publications \cite{ref_CLM_LIKKT}.
\paragraph*{Acknowledgement.---}
The works of T.A. were supported in part by Grant-in-Aid for
 Scientific Research  (No. 17K05425) from Japan Society for the Promotion of Science. Computations were carried out with KEKCC and NTUA Het Cluster. This work was also supported by computational time granted by the Greek Research and Technology Network (GRNET) in the National HPC facility --- ARIS --- under project ID ``IKKT10D''.

\end{document}